# Pulling and lifting macroscopic objects by light

Gui-hua Chen[1,2], Mu-ying Wu[1], and Yong-qing Li[1,2*]

[1]*School of Electronic Engineering & Intelligentization, Dongguan University of Technology, Dongguan, Guangdong, P.R. China*
[2]*Department of Physics, East Carolina University, Greenville, North Carolina 27858-4353, USA*

Laser has become a powerful tool to manipulate micro-particles and atoms by radiation pressure force or photophoretic force, but optical manipulation is less noticeable for large objects. Optically-induced negative forces have been proposed and demonstrated to pull microscopic objects for a long distance, but are hardly seen for macroscopic objects. Here, we report the direct observation of unusual light-induced attractive forces that allow pulling and lifting centimeter-sized light-absorbing objects off the ground by a light beam. This negative force is based on the radiometric effect on a curved vane and its magnitude and temporal responses are directly measured with a pendulum. This large force (~4.4 μN) allows overcoming the gravitational force and rotating a motor with four-curved vanes (up to 600 rpm). Optical pulling of macroscopic objects may find nontrivial applications for solar radiation-powered near-space propulsion systems and for understanding the mechanisms of negative photophoretic forces.

In Maxwell's theory of electromagnetic waves, light carries energy and momentum, and the exchange of the energy and momentum in light-matter interaction generates optically-induced forces acting on material objects. Can a centimeter-sized or larger object that can be viewed by naked eye be lifted up by a light beam and pulled towards the light source? Although optical forces have been widely applied for the manipulation of microscopic objects [1–6], they are usually unnoticeable on macroscopic objects because the magnitude of optical forces is generally much weaker than the gravitational force ($F_G$) of the large objects. Generally, when an object immersed in a gaseous environment is illuminated by light, two types of optically-induced forces are generated by the light-matter interaction. One is radiation pressure force ($F_{RP}$) arising from direct momentum transfer between the object and the incident light [1,2], which is in pN or nN range and is not enough to lift large objects. The other is photophoretic or radiometric force ($F_{RM}$) due to photo-heating effect, in which the photon energy of the incident light is first converted into the thermal energy of the object due to light absorption and then asymmetrical momentum transfer between the heated object and the surrounding gas molecules produces $F_{RM}$ [7,8], which can be several orders of magnitude larger than the radiation force $F_{RP}$ [9,10]. Optically-induced negative forces have been demonstrated to pull microscopic objects for a long distance [11,12]. The generation of optically-induced forces on macroscopic objects could trigger a new era of optomechanical applications.

Surprisingly, only a few experiments reported on the observation of optically-induced forces on centimeter-sized objects [13–16], involving force magnitude of up to 1–10$^2$ nN which was three or more orders of magnitude less than the gravity of the target objects. For examples, Magallanes et al recently observed a lateral optical force of ~1 nN on a birefringence object illuminated with a 1-W incident light, causing a lateral displacement of 0.5 mm in ~100 s [13]. Wolfe et al reported a light-induced transverse force of ~400 nN acting on a horizontal vane radiometer due to thermal creep effect [14]. Crookes demonstrated a radiometric force on black-white flat vanes or cup-shaped vanes when illuminated by light [15,16], but the magnitude and dynamic properties of Crookes radiometric forces have never been directly measured [17]. Although these forces may cause the horizontal motion or rotation of the objects, they are insufficient to manipulate the objects in vertical direction.

Here we report a direct observation of a negative optically-induced force on centimeter-sized light-absorbing objects on the order of μN that allows lifting up the object over a longitudinal distance of >5 mm in ~40 ms towards the light source by a light beam of 1-W power. This indicates that the optically-induced force acting on macroscopic objects could be larger than the object's weight by changing the surface geometry. Thus, our experiment increases the force per object's weight by three orders of magnitude compared to previous works. Our experiment relies on radiometric force $F_{RM}$ that acts on a thin vane with a curved surface structure immersed in a rarefied gas, on which a layer of black absorber is coated to increase the absorption. $F_{RM}$ could be classified into two types based on the variation in surface temperature or in heat exchange coefficient of the vane, like photophoretic forces acting on micro-particles [7,8]. The first type of radiometric force ($F_{\Delta T}$) is caused by the temperature difference between two sides of the vane, pointing from the hot to the cold side and being repulsive from the illumination light. The second type of radiometric force ($F_{\Delta \alpha}$) is caused by the unbalanced heat exchange between two sides of the vane with the gas molecules even if two sides are heated at the same temperature, due to the difference in surface geometry or accommodation coefficient. The $F_{\Delta \alpha}$-force could be attractive or repulsive in the light propagation direction [7–9]. Since Crookes' pioneering work in 1870s [15,16], extensive efforts have been made to understand the mechanisms of radiometric forces including Maxwell, Reynolds, and Einstein's work [18–20], but almost all the efforts were focused on the $F_{\Delta T}$-force acting on flat vanes that requires a temperature difference across or along the vane [14,21–23]. There are still some confusions about the dominant mechanisms in various regimes and configurations [16]. Recent applications in near-space propulsion systems driven by solar radiation [24–26], micromachines [13,21], and computational techniques [27–29] renewed the interest in $F_{RM}$. Our experiment deals with the $F_{\Delta \alpha}$-force acting on a curved vane, which doesn't require a temperature gradient on the vane and is seldom considered previously.

The experimental setup is described in detail in the supplementary materials (SM). In FIG. 1, we show direct observation of a light-induced pulling or pushing force acting on a thin absorbing curved metal vane. To determine the magnitude and direction of the radiometric force $F_{RM}$, a single cylindrical vane is suspended with an ultrafine copper wire as a pendulum in a vacuum chamber and illuminated by a light beam (see SM). The photon energy of the illumination light is absorbed by Black 2.0 paint on one side and transferred by thermal conduction to metal vane, resulting in both sides of the vane heated to an identical temperature since the used aluminum or gold vanes are excellent conductors. Therefore, $F_{RM}$ acting on a curved metal vane is the second type of radiometric force caused by unbalanced heat exchange due to the curved surface geometry. The magnitude of $F_{RM}$ is determined by $F_{RM}=mg\sin\theta$, where $m$ is the mass of the vane, $g$ is the gravitation acceleration, and $\theta$ is the angle displacement. When a laser beam is incident on the concave side of the vane, it is pulled towards the light source (FIGS. 1b and 1c, and Supplementary Video S1). Once the laser is turned off, the vane returns to its equilibrium position. However, when the laser beam is incident on the convex side, the vane is pushed along the light propagation direction (FIG. 1e). Figure 1d shows the dependence of the force magnitude on the gas pressure with a constant laser power. The maximum angle displacement is 0.172 rad at 0.1~0.2 Torr corresponding to Knudsen number Kn~0.1, and thus the maximum pulling force is ~4.4 µN from the known density and volume of the vane, where $Kn=\lambda/W$ is defined as the ratio of the molecular mean free path $\lambda$ of the surrounding gas to the characteristic length $W$ of the vane [17]. It should be noted that the radiation pressure force ($=P/c$) is ~2.3 nN, which is smaller than $F_{RM}$ by about three orders of magnitude, where $P$ is the absorbing power of incident laser, and $c$ is the speed of light.

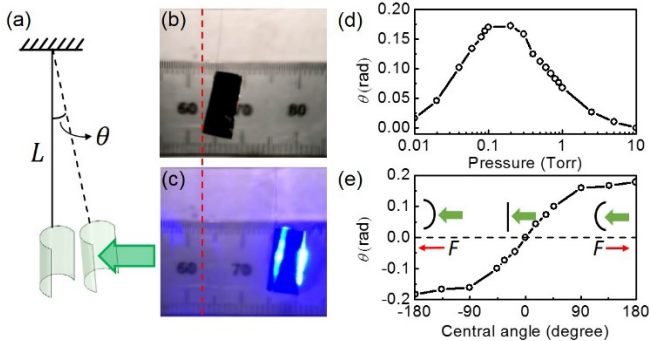

FIG. 1. Laser pulling of a concave surface. (a) Schematic setup of pulling a concave vane with a laser beam. The light-absorbing cylindrical aluminum vane (8×8 mm in size and 15 µm in thickness) is suspended with an ultrafine copper string in a pendulum and is pulled towards the laser beam. (b) Position of the black curved vane without laser illumination. (c) Position of the curved vane with laser illumination. (d) Angular displacement of the pendulum as the function of the gas pressure. (e) Angular displacement as the function of central angle of the vane at the pressure of 0.1 Torr. The laser power is 0.7 W at 450 nm.

Figure 1e shows that for a given size of the cylindrical vane, $F_{RM}$ depends on the curvature or central angle $\Theta_0$ of the cylinder surface, where $\Theta_0=W_0/R$ in radian, and $W_0$ is the arc length and $R$ is the radius. $F_{RM}$ is attractive for the illumination on concave side ($\Theta_0>0$). $F_{RM}$ is repulsive for the illumination on convex side ($\Theta_0<0$). For a flat vane ($\Theta_0=0$), $F_{RM}$ is zero so that it is not deflected by the laser beam, and this is a proof of no temperature variation between the opposite sides of the aluminum vane.

In FIG. 2, we show that the attractive radiometric force is large enough to overcome the gravitational force $F_G$ and lift an ultrathin curved gold vane off the ground by the laser illumination. $F_G$ of the curved gold leaf is estimated to be 1.85 µN from the known density and volume of the vane. The vane can be quickly pulled off from the supporting surface and lifted up until the top of the vacuum chamber in ~40 ms by the attractive $F_{RM}$, see FIG. S1 and Supplementary Video S2.

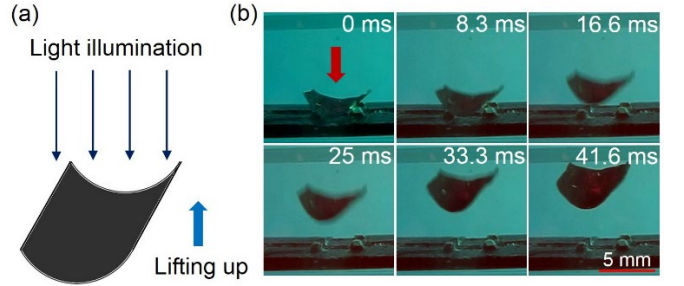

FIG. 2. Lifting of a curved vane. (a) An ultrathin curved vane is lifted when the concave side is illuminated. (b) Time-lapsed images of a cylindrical gold leaf (with 7×7 mm in size and 0.2 µm in thickness) when the laser beam is turned on. The laser power is 1 W at 650 nm and the pressure is 0.3 Torr.

We interpret the observed negative $F_{RM}$ acting on a curved vane by the combination of pressure force ($F_n$) and shear force ($F_t$), in contrast to the area force, edge force, and shear force acting on a flat vane with a temperature gradient [17,27]. When a heated vane at a temperature $T_s$ is immersed in a rarefied gas with a temperature $T_0$ ($T_0<T_s$), heat exchange between the vane and surrounding molecules occurs via molecular collisions and convectional gas flow (FIG. 3). The pressure force (like the area force and edge force) is caused by the difference in momentum flux between the gas molecules leaving the two sides of the vane and acts in the normal direction to the surface [27]. The shear force is in the tangential direction of the surface and is caused by the interaction of thermal creep flow due to Reynolds effect [19], since the force on the gas is equal and opposite to the force on the surface. It has been shown that the gas temperature near the edge is lower than that near the center of a hot vane, so that a gas flow parallel to the surface moves from the edge to the center region [14,28]. If the vane without a temperature gradient is flat (FIG. 3a), the net $F_{RM}$ is zero because $F_n$ and $F_t$ are cancelled due to the balance in heat exchange between two sides [27,28]. However, if the vane is curved and the concave side is illuminated (FIG. 3b), a non-zero $F_{RM}$ is generated in the upward direction. As shown in the Supplementary Materials, $F_n$ acting on a curved vane is generated by the unbalanced momentum flux between the molecules leaving the concave and convex sides of the vane and acts in the direction towards the concave side, because the molecules incident on convex side only experience single reflection and the molecules incident on the concave side may experience multiple reflections due to the geometry effect. In free-molecular regime (Kn>10 at low gas pressure), the pressure force is the dominant mechanism of $F_{RM}$ [18].

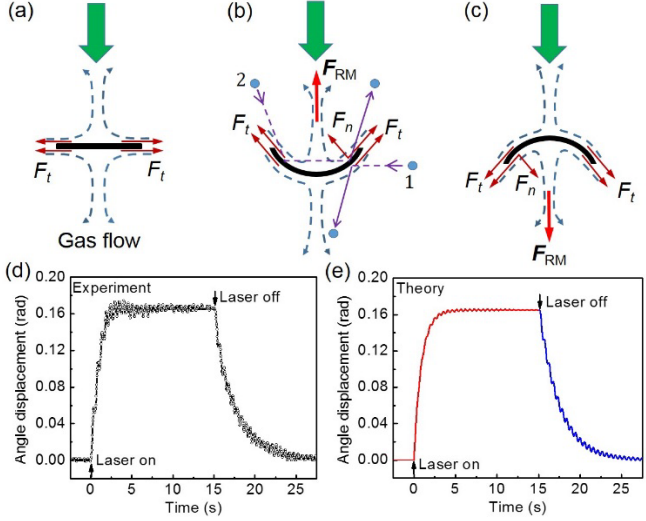

FIG. 3. Radiometric force $F_{RM}$ on a heated curved vane by molecular collisions. (a) Schematic showing tangential shear forces on the surface of a hot flat vane due to gas flow. The net $F_{RM}$ on the vane is zero due to the geometry balance. (b) Pressure force and shear force on a surface element of a concave vane and the total $F_{RM}$ pointing to the upward direction. Single collision of the incident molecules occurs on the convex side, and twice or multiple collisions of the incident molecules occur on the concave side. (c) $F_{RM}$ on a convex vane. (d) Experimental angle displacement-time graph of a pendulum (with an 8×8 mm cylindrical vane) when the laser beam of 1 W is turned on or off at 0.1-Torr pressure. (e) Theoretical modeling of dynamic angular displacement of the pendulum with $\tau_r$=0.9 s, $\tau_f$=2.4 s, $\beta$=0.12 s$^{-1}$, and $L$=70 mm.

When the pressure increases to transitional flow region (0.01<Kn<10), the molecule flux hitting on the vane increases, but the molecular mean free path is reduced so that the collisions between the incoming molecules and the reflected molecules will equilibrate the momentum difference due to the multiple reflections on the concave side. Thus, the pressure force would be optimum at Kn~0.1 and decreases in the continuum flow regime. On the other hand, in transitional flow region the shear force plays an important role since thermal creep flow of gas molecules occurs at low Kn value [14,27]. As shown in FIG. 3b, the net shear force is toward the concave side. Therefore, the integration of the pressure force and shear force over the entire surface generates a non-zero attractive $F_{RM}$ pointing to the upward direction. Similarly, if the convex side of the vane is illuminated (see FIG. 3c), $F_{RM}$ acting on it is repulsive, pointing to the downward direction.

The upper limit of the radiometric force could be estimated by $F_{RM}=P_0A$, when the pressure force on one side of the vane is zero, where $P_0$ is the gas pressure and $A$ is the surface area of the vane. This gives ~840 μN for the cylindrical vane and the pressure of 0.1 Torr in FIG. 1, which is about two orders in magnitude larger than our observed force, suggesting that the force magnitude could be further increased by optimizing the experimental parameters.

$F_{RM}$ on the curved vane depends on the temperature difference $T_s$-$T_0$ between the vane and the surrounding molecules, the geometry of the vane, the gas pressure, and the illumination light intensity. When the illumination light is turned on or off, $F_{RM}$ is changed in response to the change in temperature $T_s$ of the vane. To determine how fast $F_{RM}$ is changed temporally, we measure the dynamic motion of the vane in a pendulum. Figure 3d shows the angle displacement-time graph of the pendulum when the laser beam is turned on or off, indicating that the vane moves to the steady-state position with a quick rise time and moves back with a slower fall time and with an oscillation frequency of 1.88 Hz. The dynamic motion of the vane can be described by the pendulum equation

$$mL\ddot{\theta} = F_{RM} - F_{RP} - mg\sin\theta - \gamma L\dot{\theta}, \quad (1)$$

where $L$ is the length of the pendulum, $F_{RP}$ is the radiation pressure force that can be neglected as $F_{RP} \ll F_{RM}$, and $\gamma$ is Stokes' drag constant. In steady state after the laser beam is turned on, $F_{RM}$ reaches the maximum $F_{RM,0}=mg\sin\theta_0$, where $\theta_0$ is the steady-state deflection angle. Since $F_{RM}$ is proportional to ($T_s$-$T_0$) [7,8], which changes exponentially when the laser power is modulated [9], it can be expressed as $F_{RM}(t)=F_{RM,0}[1-\exp(-t/\tau_r)]$ or $F_{RM}(t)=F_{RM,0}\exp(-t/\tau_f)$, where $\tau_r$ is the rise time and $\tau_f$ is the fall time of the temperature of the vane when the laser beam is turned on or off, respectively. Equation (1) can be expressed as

$$\ddot{\theta} + 2\beta\dot{\theta} + \omega_0^2\theta = f_0(1 - e^{-t/\tau_r}), \quad (2)$$

or

$$\ddot{\theta} + 2\beta\dot{\theta} + \omega_0^2\theta = f_0 e^{-t/\tau_f}, \quad (3)$$

where $\beta=\gamma/(2m)$, $f_0=F_{RM,0}/(mL)$, and $\omega_0 = \sqrt{g/L}$ is the natural frequency, $\omega = \sqrt{\omega_0^2 - \beta^2}$ is the oscillation frequency of the pendulum. Figure 3e shows the theoretical modeling of dynamic angular displacement of the pendulum with $\tau_r$=0.9 s, $\tau_f$=2.4 s, and $\beta$=0.12 s$^{-1}$, where $\omega_0=2\pi\times1.884$ Hz calculated from the length $L$=70 mm. This result indicates that both the radiometric force and the temperature of the vane increase with a short rise time of ~1 s when the laser is turned on and decrease with a slightly slower time when the laser is off.

Figure 4 shows that $F_{RM}$ can be used to drive the rotation of a motor with four-curved vanes. When the concave side of the cylindrical aluminum vane is illuminated by a laser beam, the attractive $F_{RM}$ pulls the vane towards the laser beam (Supplementary Video S3). When the convex side of the vane is illuminated, the repulsive $F_{RM}$ turns the vane away from the beam, in the same rotation direction as the Crookes radiometer with black-white flat vanes (FIG. 4b). When the vane is flat, no rotation is observed, indicating no temperature difference between two sides of the vane. Figure 4f shows the rotation speed of Crookes radiometer with concave, flat-plate, and convex aluminum vanes of the same size as the function of the laser power. Figure S3 shows the dependence of rotation speed of a motor with smaller cylindrical vanes (10×10 mm) on central angle and gas pressure, with a speed up to 600 rpm. It should be noted that the difference between our experiment and previous experiments [15,21] is that the spot size of the illumination light is smaller than the size of single curved vanes so that only one vane is illuminated at any time during the rotation. This excludes the complicated procedures to determine the force direction on cup-shaped vanes in Crookes experiment with a candle illumination [15], and excludes the possibility of the asymmetric heating that could result in a temperature gradient between two sides of the vanes in a micromotor [21].

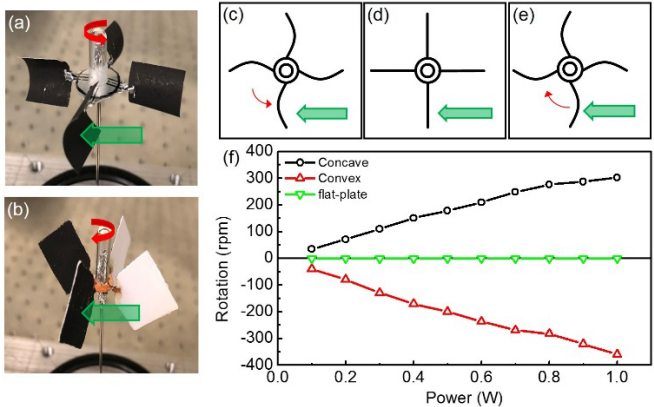

FIG. 4. Crookes radiometer with four-curved vanes or black-white flat vanes. (a) When a concave absorbing aluminum surface is illuminated by a laser beam, the pulling force turns the vane towards the beam. (b) When a black-white flat vane is illuminated, the pushing force turns the vane away from the beam. (c)-(e) Laser-induced attractive, zero, or repulsive force on a concave (c), flat-plate (d), or convex surface (e). (f) Rotation speed of Crookes radiometer with concave, flat-plate, and convex aluminum vanes of the same size (16×16×0.05 mm) verse the laser power. The pressure is 0.02 Torr.

Radiometric force $F_{RM}$ on macroscopic objects might be applied for near-space propulsion systems in low pressure environment [24–26]. It has been proposed that an array of multiple planar vanes with a thermal insulator between two surfaces powered by solar radiation could be employed in radiometric-based propulsion systems for a vehicle operating at an altitude from 40 to 80 km. However, this first-type $F_{RM}$ is repulsive along solar radiation direction so that the vehicle is generally pushed in one direction. If the planar multiple vanes are replaced by multiple curved vanes, the generated force can be attractive or repulsive. Thus, the propulsion direction of the vehicle could be controlled by turning the angle of the curved vanes relative to the solar radiation direction. Also, the origin of the change in the accommodation coefficient over the surface of tiny micro-particles is difficult to quantitatively characterize [9,12]. The effect of surface geometry of the macroscopic vanes on the second type $F_{RM}$ might provide an approach to the understanding of the mechanisms on the negative photophoretic force.

In summary, we measure the magnitude and dynamic properties of light-induced attractive and repulsive radiometric forces on macroscopic curved metal vanes. By changing the surface geometry, we demonstrate that the attractive radiometric force can be greater than the gravitational force and lift a centimeter-sized gold surface by a light beam. This force can also drive the rotation of a motor with four-curved vanes with a high speed. Light-induced attractive forces on macroscopic objects may find potential applications for radiometric-based near-space propulsion systems.

This work was supported by the National Natural Science Foundation of China (Grant No. 61775036) and by the high-level talents program of Dongguan University of Technology (Grant. No. KCYCXPT2017003). *Corresponding author. liy@ecu.edu

**Supplement materials:**
FIG. S1 Lifting up of a cylindrical gold leaf. Time-lapsed images of a cylindrical gold leaf when illuminated by a laser power under the pressure of 0.3 Torr.
FIG. S2 Schematic showing the tangential shear force and pressure force on a hot curved vane.
FIG. S3 Dependence of rotation speed of a Crookes radiometer with four curved vanes on central angle and gas pressure.
Video S1 Observation of the deflection and restoration of a suspended curved vane when the laser beam is turned on and off.
Video S2 Observation of the lifting of a curved gold leaf by an attractive radiometric force when a laser beam is incident on the vane under the pressure of 0.3 Torr.
Video S3 Observation of the rotation of a Crookes radiometer with four-curved vanes with a rotation speed up to 600 rpm driven by an attractive radiometric force when a laser beam is incident on the vane.

**DATA AVAILABILITY**
The data that support the findings of this study are available from the corresponding author upon reasonable request.

# Pulling and lifting macroscopic objects by light

Gui-hua Chen[1,2], Mu-ying Wu[1], and Yong-qing Li[1,2*]

[1] School of Electronic Engineering & Intelligentization, Dongguan University of Technology, Guangdong 523808, P.R. China.
[2] Department of Physics, East Carolina University, Greenville, North Carolina 27858-4353, USA. *e-mail: liy@ecu.edu

# Supplementary Information

## Materials and Methods

**Laser pulling and pushing of a cylindrical metal vane:** The absorbing cylindrical vane was made of a thin aluminum sheet (8×8 mm in size and 15 μm in thickness), which was painted with Black 2.0 paint on both sides and bended as the cylindrical shape with a cylindrical rod of a specific radius. The cylindrical aluminum vane was suspended with an ultra-fine copper wire (70 mm in length and 10 μm in diameter) to form a pendulum inside a vacuum jar chamber (15 cm in height and 12 cm in diameter), see FIG. 1a. The pressure of the vacuum chamber was controlled with a Turbo pump station and a set of valves, which were closed off during the experiment such that there was no air flow inside the chamber while the pressure in the chamber was maintained at constant. A continuous-wave collimated laser beam at 450 nm (with a beam size of ~2×4 mm and the maximum power of 1.2 W) was incident on the concave or convex side of the absorbing cylindrical vane. The temperature of the absorbing surface was measured to be ~50 ºC higher than that of the vacuum chamber with an infrared thermometer. Upon the illumination of the laser beam, the absorbing aluminum vane was deflected from its equilibrium position, which was recorded with a video camera. The deflection angle was determined by the ratio of the displacement and the string length, and the mass (and gravitational force) of the vane was determined by the known density and volume of the aluminum vane. See Online Supplementary Video S1.

**Laser lifting of a cylindrical metal vane:** An ultrathin gold foil leaf (7×7 mm in size and ~0.2 μm in thickness) was painted with Black 2.0 paint and gently bended as the cylindrical surface. The cylindrical gold leaf was placed on the top of a flat surface with four contact points and with the concave side facing up (see FIG. 2b). The pressure of the vacuum chamber (a 10×10×50 mm quartz cuvette cell or a large jar chamber) was controlled with a Turbo pump station. When the illuminating laser beam of 1 W at 650 nm was turned on, the absorbing gold leaf was rapidly lifted up, which was recorded with a slow-motion video camera (240 fps), see Online Supplementary Video S2. A green short-pass filter was used in the front of the camera to block the scattered laser light.

**Laser pulling of Crookes radiometer with four-curved vanes:** The black cylindrical curved vane was made with matte black aluminum sheet (16×16 mm in size and 50 μm in thickness, Cinefoil, Rosco Corp.). Four curved vanes were mounted on a small glass tube (see FIG. 4a), which was placed on the top a stainless-steel needle to allow for rotation in a vacuum chamber with the minimized resistance. A collimated laser beam at 650 nm (with a beam size of ~5×5 mm) was incident on the concave or convex side of one curved vane and drive the rotation of the Crookes radiometer. A video camera was used to record the rotation with a 550-nm short-pass filter inserted in the front of the camera to block the laser scattering light, see Online Supplementary Video S3. A photodiode was used to detect the attenuated laser beam that passed through the vanes and the rotation speed was determined from the periodic photodiode signal.

## Pressure force acting on a curved vane

The pressure force $F_n$ acting on a curved vane due to the collision of gas molecules is generated by the unbalanced momentum flux between the molecules leaving the concave and convex sides of the vane and acts in the direction towards the concave side [see Figs. S2(b) and S2(c)]. As shown in Fig. S2(b), in free-molecular regime (Kn>10 at low gas pressure), all molecules hitting on a surface element $\Delta S$ from the convex side (such as molecule 1) only experience single reflection on the surface, leaving at a higher temperature with the difference of $T-T_0=\alpha(T_s-T_0)$, where $\alpha$ is the accommodation coefficient of the surface which is assumed to be uniform on the whole vane. However, some molecules hitting on $\Delta S$ from the concave side (such as molecule 2) experience two or multiple reflections on the surface, thus leaving with a temperature difference of $T-T_1=\alpha(T_s-T_0)-\alpha^2(T_s-T_0)$, where $T_1=T_0+\alpha(T_s-T_0)$ is the molecule's temperature before hitting on Δ$S$ due to the first reflection. As a result, the energy difference between a pair of molecules (e.g. molecules 1 and 2) leaving Δ$S$ is Δ$E=\alpha^2 k(T_s-T_0)$, leading to a net momentum change pointing from the convex to the concave sides, where $k$ is Boltzmann constant. Note that the molecule flux hitting on Δ$S$ from both sides should be identical because $F_n$ should be zero if there is no temperature difference between the vane and surrounding molecules.

## Supplementary Movies
**Supplementary Video 1**
Macroscopic observation of the deflection and restoration of a suspended curved vane when the laser beam is turned on and off.
**Supplementary Video 2**
Macroscopic observation of the lifting of a curved gold leaf by an attractive radiometric force when a laser beam at 650 nm with an optical power of 1 W is incident on the vane under the pressure of 0.3 Torr.
**Supplementary Video 3**
Macroscopic observation of the rotation of a Crookes radiometer with four-curved vanes driven by an attractive radiometric force when a laser beam at 450 nm with an optical power of 0.8 W is incident on the vane under the pressure of 0.02 Torr. When the laser beam is turned on, the radiometer starts the rotation and reaches the maximum speed in about 1 min. When the laser beam is turned off, the radiometer decelerates and stops in about 1 min.

## Supplementary Figures
**Figure S1. Lifting up of a cylindrical gold leaf (7×7 mm in size, 0.2 μm in thickness) with 1 W-laser at 650 nm**

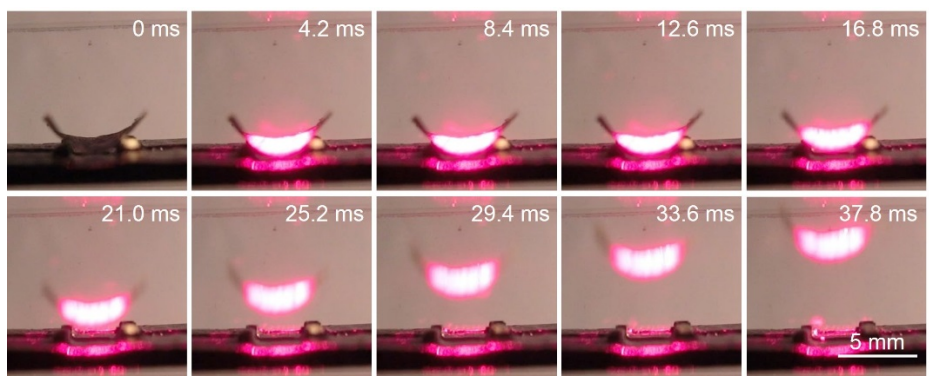

**FIG. S1. Lifting up of a cylindrical gold leaf**. Time-lapsed images of a cylindrical gold leaf (7x7 mm in size, 0.2 μm in thickness) when illuminated by a laser power at 650 nm with an optical power of 1 W under the pressure of 0.3 Torr.

**Figure S2. Schematic showing the tangential shear force and pressure force on a hot curved vane**

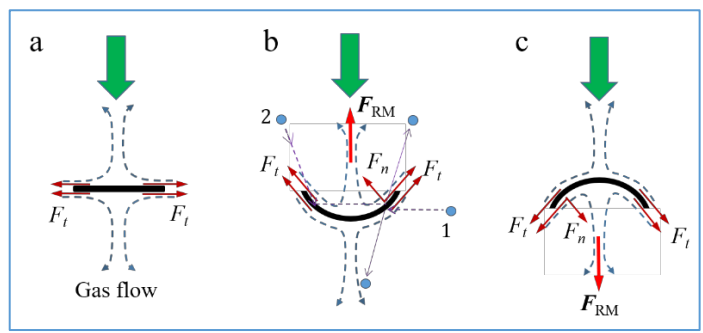

**Fig. S2. Schematic showing the tangential shear force and pressure force on a hot curved vane**. (a) Schematic showing tangential shear forces on the surface of a hot flat vane due to gas flow. The net $F_{RM}$ on the vane is zero due to the geometry balance. (b) Pressure force and shear force on a surface element of a concave vane and the total $F_{RM}$ pointing to the upward direction. Single collision of the incident molecules occurs on the convex side, and twice or multiple collisions of the incident molecules occur on the concave side. (c) Pressure force and shear force on a surface element of a convex vane and the total $F_{RM}$ pointing to the downward direction.

**Figure S3. Dependence of the rotation speed of a Crookes radiometer with four curved vanes on central angle and gas pressure**

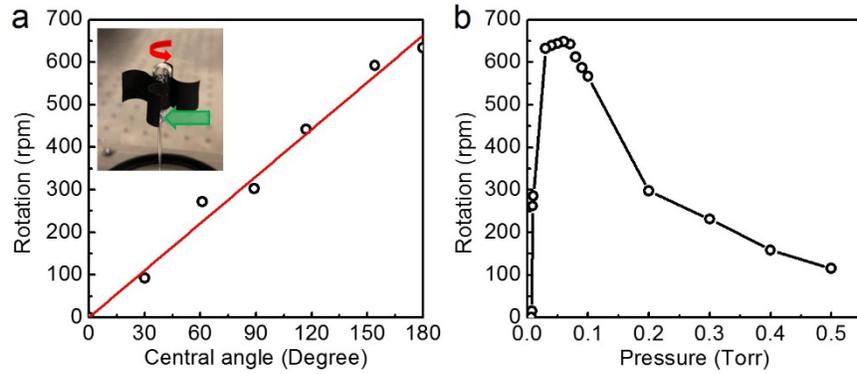

**FIG. S3. Dependence of rotation speed of a Crookes radiometer with four curved vanes on central angle and gas pressure.** (a) Dependence of rotation speed on central angle $\Theta_0$ of the curved-vane (10×10 mm in size and 50 μm in thickness, shown in the insert). The laser power is 1 W and the pressure is 0.03 Torr. The rotation speed is proportional to the central angle of the vanes (fit in line). (b) Dependence of rotation speed on the pressure with the central angle $\Theta_0=\pi$.